\documentstyle[aps,prd,preprint,epsfig]{revtex}
\tighten
\begin{document}
\preprint{
\parbox{45mm}{
\baselineskip=12pt
YCTP-P14-98\\
July, 1998
\vspace*{1cm}}}
\title{Quark Mass Hierarchy\\
       in the Supersymmetric Composite Model}
\author{Noriaki Kitazawa\thanks{e-mail: kitazawa@zen.physics.yale.edu}}
\address{Department of Physics, Yale University,
         New Haven, CT 06520, USA
         \thanks{On leave: Department of Physics,
                           Tokyo Metropolitan University,
                           Tokyo 192-03, Japan}}
\maketitle
\begin{abstract}
A mechanism to have the quark mass hierarchy
 in the supersymmetric composite model is proposed.
The source of the hierarchy
 is the kinetic-term mixing between composite quarks.
Such mixing can be expected, if quarks are composite particles.
A model in which the mechanism works is constructed,
 although it is not perfectly realistic
 because of some assumptions and unnaturalness.
\end{abstract}
\newpage

\section{Introduction}
\label{sec:intro}

Solving the quark and lepton mass hierarchy problem
 is one of the most important subject
 in the elementary particle physics.
Many mechanisms for the mass hierarchy have already been proposed.
For example,
 in the $SU(5)$ grand unified theory
 the mass splitting between the bottom quark and the tau lepton
 can naturally be explained by the unification of their Yukawa couplings
 and scale dependence governed by the renormalization group equation.
In the extended technicolor theory
 the generation of the mass hierarchy is expected
 as a result of the dynamics of the strong coupling chiral gauge theory.

Considering the compositeness of quarks and leptons
 is one of the attempt to solve the mass hierarchy problem.
Especially the supersymmetric composite model is attractive,
 since we can naturally have tightly-bounded states
 whose masses are much smaller than their compositeness scales.
This fact is suggested by experiments.
Several interesting mechanisms
 for the mass hierarchy in the supersymmetric composite model
 have already been proposed.
Having the mass hierarchy
 by the interplay between the chiral symmetry and the supersymmetry
 is proposed in Refs.\cite{BPY,BMV}.
The power of the ratio of different energy scales
 can give the mass hierarchy\cite{FN,Strassler,KLS,Hayakawa}.

Almost all recent attempts
 to have the mass hierarchy in the supersymmetric composite model
 are based on the existence of hierarchical energy scales.
Many models are constructed
 in which hierarchical Yukawa couplings are generated
 as the ratios of different energy scales.
In these models
 we usually have to prepare many different energy scales
 and assume some non-renormalizable interactions
 in the Lagrangian at tree level.
In this paper we propose a mechanism
 which is not based on the hierarchical energy scales.

\begin{figure}
$$
\mbox{\epsfig{file=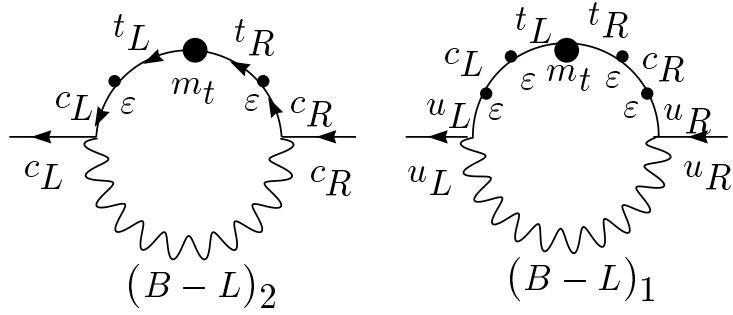,width=10cm}}
$$
\caption{Diagrams for the charm and down quark mass generation
         through the kinetic-term mixing and $B-L$ interactions.}
\label{B-L}
\end{figure}

The basic idea is simple.
We assume the kinetic-term mixing among quarks like
\begin{equation}
 {\cal L}_{kin} =
  \left(
   \begin{array}{ccc}
    {\bar u} & {\bar c} & {\bar t}
   \end{array}
  \right)
  \left(
   \begin{array}{ccc}
    1 & \varepsilon & 0 \\
    \varepsilon & 1 & \varepsilon \\
    0 & \varepsilon & 1
   \end{array}
  \right)
  i D^\mu \gamma_\mu
  \left(
   \begin{array}{c}
    u \\ c \\ t
   \end{array}
  \right).
\end{equation}
The large mixing,
 $\varepsilon \simeq 0.1$, can be obtained,
 if quarks are composite particles\cite{Kitazawa},
 and the texture of the mixing can be understood
 as a result of the dynamics of compositeness.
This kinetic-term mixing
 is a source of the generation symmetry violation.
Furthermore,
 we assume the interactions which are peculiar to each generation.
For example, to be concrete,
 we consider three gauged $U(1)_{B-L}$
 interactions in each generation,
 where $B$ and $L$ denote the baryon and lepton numbers, respectively.
To be realistic, these interactions must be spontaneously broken,
 but we do not assume any hierarchical structure
 in the energy scales of the breaking and gauge couplings.
If the mass of the top quark is generated,
 the masses of the charm and up quarks are generated
 thorough the diagrams of Fig.\ref{B-L}.
These masses are obtained as
\begin{equation}
 m_c \simeq {{g^2} \over {16 \pi^2}} \varepsilon^2 m_t,
\qquad
 m_u \simeq {{g^2} \over {16 \pi^2}} \varepsilon^4 m_t
\end{equation}
 up to the common logarithmic correction,
 where $g$ is the gauge coupling.
In case of the strong coupling, $g^2/16 \pi^2 \simeq 1$,
 we can have a good hierarchy
\begin{equation}
 {{m_c} \over {m_t}} \simeq {{m_u} \over {m_c}} \simeq \varepsilon^2
\end{equation}
 with $\varepsilon \simeq 0.07$.
Here, the masses of up-type quarks at $1$ GeV are
 $m_t \simeq 190$ GeV, $m_c \simeq 1.4$ GeV and $m_u \simeq 5$ MeV.
If $U(1)_{B-L}$ gauge interactions
 are unique and common to each generation,
 the one-loop contribution gives
 just a correction to the kinetic-term mixing
 and does not generate mass eigenvalues.

In the following
 we construct a model in which the above situation
 (but not the gauged $U(1)_{B-L}$) is realized
 to show that the mechanism is possible.
Although the model is not completely realistic
 because of some assumptions and unnaturalness,
 it is worth developing.

In the next section
 the one-generation (the third generation) model
 with dynamical supersymmetry breaking
 is described.
The model is based on the works of Refs.\cite{NS,KO}.
In section \ref{sec:three}
 the model is extended to the three-generation model
 by including other two generations.
The dynamics of the compositeness in each three generation are the same,
 namely the same structure is repeated three times.
But the compositeness scales of each generation
 are related with each other,
 since confining forces
 are expected to be ``unified'' at a certain scale.
The discussion on the origin of the kinetic-term mixing
 is also given in this section.
In section \ref{sec:hierarchy}
 it is explained how the quark mass hierarchy is realized in the model.
Both the up-type and down-type quarks are discussed.
We see that the mechanism which is briefly sketched above
 actually works in the model.
In the last section we conclude by listing up
 both the satisfactory and unsatisfactory points of the model.

\section{One-Generation Model with Supersymmetry Breaking}
\label{sec:one}

The model of the third generation is constructed in this section.
The difference of the third generation from other two generations
 is that it strongly couples to the supersymmetry breaking sector.
The model described in this section
 is essentially the same of that is proposed in Ref.\cite{KO},
 except for the introduction of the right-handed neutrino
 and the difference of the notation.

\subsection{Particle contents and Interactions}
\label{subsec:contents}

Two new gauge interactions are introduced
 in addition to the gauge interactions of the standard model
 $SU(3)_C \times SU(2)_L \times U(1)_Y$.
The $SU(2)_3$ hypercolor is introduced
 as the confining force to have composite particles
 in the third generation,
 and the $SU(2)_S$ supercolor is introduced
 to realize the dynamical supersymmetry breaking.
The fundamental particles of the third generation
 are shown in Table \ref{contents-3},
 where $U(1)_{B-L}$ is a global symmetry.
All the particles in Table \ref{contents-3}
 are singlet under the $SU(2)_S$ supercolor.
In the following
 all fields are superfields without special notice.
Note that
 there is no particle
 which belongs to the high-dimensional representation of gauge groups,
 and the representation is vector-like
 under each factor of the gauge group.
In this sense, the particle contents are simple.
We mention that
 the $U(1)_{B-L}$ symmetry can be gauged
 because of the inclusion of the right-handed neutrino,
 although we do not gauge it.
The renormalizable tree-level superpotential
 which is general with respect to the symmetry is
\begin{equation}
 W_H = \eta {\bar H}_3 [D_3 N_3] + \eta_{LQ} {\bar \Phi}_3 [C_3 N_3],
\label{WH-3}
\end{equation}
 where square brackets denote the contraction of $SU(2)_3$ indices,
 $[DN] \equiv \epsilon_{\alpha\beta} D^\alpha N^\beta$, for example.

The particle contents of the supersymmetry breaking sector
 are shown in Table \ref{susy-br}
 \footnote{The explanation of the symmetry $U(1)_i$
           will be given in section \ref{sec:three}}.
All the particles in Table \ref{susy-br}
 are singlet under the standard model gauge group.
Since some particles have the $SU(2)_3$ hypercolor,
 the supersymmetry breaking due to the $SU(2)_S$ supercolor dynamics
 is transmitted to the superpartners of known particles.
If we neglect the $SU(2)_3$ hypercolor,
 the system has $SU(4)$ global symmetry
 ($({\tilde Q} \ Q) \in 4$ and $({\tilde Z} \ Z' \ Z) \in 6$).
But we preserve only its subgroup $Sp(2) \sim SU(2) \times SU(2)$,
 and one of the $SU(2)$ subgroups of $Sp(2)$ is gauged as $SU(2)_3$.
Therefore the global symmetry of this sector is
 $U(1)_R \times U(1)_{B-L} \times SU(2)_{global}
  \times Sp(2)/(SU(2)_{global} \times SU(2)_3)$.
The renormalizable tree-level superpotential
 which is general with respect to the symmetry is
\begin{equation}
 W_S = \lambda {\rm tr}( {\hat Z} {\hat {\cal V}} )
     + 4 \lambda_V Z {\cal V}
     + M_X [X_1 X_2],
\label{WS}
\end{equation}
 where
\begin{eqnarray}
 {\hat {\cal V}} &\equiv& \epsilon^{ab}
 \left(
  \begin{array}{c}
   {\tilde Q} \\ Q 
  \end{array}
 \right)_a
 \left(
  \begin{array}{cc}
   {\tilde Q} & Q \\
  \end{array}
 \right)_b
\nonumber\\
 &=&
 \left(
  \begin{array}{cccc}
   0 & {\cal V}+{\cal V}' & {\tilde {\cal V}}_{11} &
   {\tilde{\cal V}}_{21} \\
   -({\cal V}+{\cal V}') & 0 & {\tilde {\cal V}}_{12} &
   {\tilde{\cal V}}_{22} \\
   -{\tilde {\cal V}}_{11} & -{\tilde {\cal V}}_{12} &
   0 & {\cal V}-{\cal V}' \\
   -{\tilde {\cal V}}_{21} & -{\tilde {\cal V}}_{22} &
   -({\cal V}-{\cal V}') & 0 
  \end{array}
 \right),
\label{def-calV}
\\
 {\hat Z} &\equiv&
 \left(
  \begin{array}{cccc}
   0 & Z+Z' & {\tilde Z}^{11} & {\tilde Z}^{21} \\
   -(Z+Z') & 0 & {\tilde Z}^{12} & {\tilde Z}^{22} \\
   -{\tilde Z}^{11} & -{\tilde Z}^{12} & 0 & Z-Z' \\
   -{\tilde Z}^{21} & -{\tilde Z}^{22} & -(Z-Z') & 0 
  \end{array}
 \right).
\end{eqnarray}
Here, $a$ and $b$ are $SU(2)_S$ indices,
 and the indices of fields ${\tilde {\cal V}}$ and ${\tilde Z}$
 are explicitly written as
 ${\tilde {\cal V}}_{\alpha i}$ and ${\tilde Z}^{\alpha i}$
 with the $SU(2)_3$ index $\alpha$
 and the $SU(2)_{global}$ index $i$.
The mass of $X_1$ and $X_2$ is introduced by hand, for simplicity,
 and the origin is not discussed in this paper.
We simply expect that
 the order of the mass $M_X$
 is the energy scale of the $SU(2)_S$ dynamics.
It is possible to have such mass,
 if we introduce additional fields\cite{Okada}.

We introduce the tree-level superpotential
 which connects above two sectors.
\begin{eqnarray}
 W_d &=& \kappa_{{\bar H}3} {\bar H}_3 [D_3 X_1]
       + \kappa_{l3} l_3 [D_3 X_2]
\nonumber\\
     &+& \kappa_{{\bar \Phi}3} {\bar \Phi}_3 [C_3 X_1]
       + \kappa_{{\bar d}3} {\bar d}_3 [C_3 X_2]
\nonumber\\
     &+& \kappa_{{\bar \nu}3} {\bar \nu}_3 [N_3 X_1].
\label{Wd-3}
\end{eqnarray}
This renormalizable superpotential
 is general with respect to the symmetry.
We can determine anomaly-free $U(1)_R$ charge assignment
 on third generation particles
 so that whole superpotential,
 Eqs.(\ref{WH-3}),(\ref{WS}) and (\ref{Wd-3}), has charge two.
Though the result is not unique, the assignment
\begin{equation}
 \begin{array}{ccccccccc}
 & C_3 & D_3 & N_3 &
   {\bar d}_3 & l_3 & {\bar \nu}_3 &
   {\bar \Phi}_3 & {\bar H}_3 \\
 U(1)_R & 1/6 & 1/4 & 1 & 5/6 & 3/4 & 0 & 5/6 & 3/4 
 \end{array}
\end{equation}
 is possible, for example.

This $U(1)_R$ symmetry has to be spontaneously broken
 to have the masses of gauginos in the standard model.
Since the resultant Nambu-Goldstone boson
 can couple to quarks and leptons,
 it may cause astrophysical and cosmological problems
 (the R-axion problem).
Therefore,
 the $U(1)_R$ symmetry has to be explicitly broken
 by the tree-level superpotential, gauge anomaly and so on.
If we believe the supergravity,
 the $U(1)_R$ symmetry is necessarily broken
 to have small cosmological constant,
 and the R-axion problem is solved\cite{BPR}.

\subsection{Generating the One-Generation Structure}
\label{subsec:confinement}

In this section we describe
 how the one-generation structure is realized at low energy
 after the confinement of the $SU(2)_3$ hypercolor.

The $SU(2)_3$ hypercolor interaction
 becomes strong at the confinement scale $\Lambda_3$
 where all the standard model interactions are weak.
It is enough to consider only the particles in Table \ref{contents-3},
 since it will be shown in the next section that
 all $SU(2)_3$ charged particles in Table \ref{susy-br}
 decouple above the scale $\Lambda_3$.
It is expected that
 the confinement without chiral symmetry breaking occurs
 in the system of supersymmetric $SU(2)$ gauge theory
 with six doublets matter ($N_f=3$)\cite{Seiberg}.
Namely, the massless low energy effective fields
\begin{eqnarray}
 M &\sim& \epsilon^{\alpha\beta}
          \left(
           \begin{array}{c}
            C_3 \\ D_3 \\ N_3
           \end{array}
          \right)_\alpha
          \left(
           \begin{array}{ccc}
            C_3 & D_3 & N_3
           \end{array}
          \right)_\beta
\nonumber\\
    &=&   \left(
           \begin{array}{ccc}
            \left[C_3 C_3\right] & \left[C_3 D_3\right] &
             \left[C_3 N_3\right] \\
            \left[D_3 C_3\right] & \left[D_3 D_3\right] &
             \left[D_3 N_3\right] \\
            \left[N_3 C_3\right] & \left[N_3 D_3\right] & 0
           \end{array}
          \right)
\label{eff-field}
\end{eqnarray}
 are expected in our model.
These composite particles
 are identified to the third generation particles as
\begin{equation}
 M = \left(
      \begin{array}{ccc}
       {\bar u}_3 & q_3 & \Phi_3 \\
       -q_3^T & {\bar e}_3 & H_3 \\
       -\Phi_3^T & -H_3^T & 0 
      \end{array}
     \right),
\label{identify}
\end{equation}
 where ${\bar u}_3$ and $q_3$ are the right-handed top quark
 and the left-handed weak doublet quark in the third generation,
 respectively, 
 ${\bar e}_3$ is the right-handed tau lepton,
 $H_3$ is the weak doublet Higgs which can couple to up-type quarks,
 and $\Phi_3$ is the leptoquark.
The particle contents of the third generation
 with a pair of Higgs and a pair of leptoquarks
 are realized by these composite particles
 and the particles in Table \ref{contents-3}
 which do not have the $SU(2)_3$ hypercolor.
Furthermore,
 it is expected that the superpotential
\begin{eqnarray}
 W_{dyn}
  &=& - \alpha {\rm Pf} M
\nonumber\\
  &=& \alpha \left(
              H_3 q_3 {\bar u}_3
              + \Phi_3 q_3 q_3
              + \Phi_3 {\bar u}_3 {\bar e}_3
             \right)
\label{Wdyn-3}
\end{eqnarray}
 is dynamically generated,
 where $\alpha$ is a dimensionless constant.
The first term of this superpotential
 can be the Yukawa coupling for the top quark mass.
The top quark naturally becomes massive,
 once the electroweak symmetry is spontaneously broken
 by the vacuum expectation value of $H_3$,
 since there is no global chiral symmetry
 which forbids the mass of the top quark in this model.

According to the naive dimensional analysis\cite{NDA},
 the coefficient $\alpha$ is of the order of $4\pi$
 for canonically normalized composite fields.
This is too large for the top-quark Yukawa coupling
 (the value of $\tan \beta$ have to be very small).
If we naively consider the renormalization group running of $\alpha$
 from the compositeness scale to the low energy,
 it can be realized $\alpha \simeq 1$ at the low energy.
But considering the renormalization group
 is inconsistent with the fact that
 the dynamically generated superpotential
 is a part of the exact effective action in the low energy limit.
If we evaluate the quantum correction
 using the exact K\"ahler potential,
 the result must give only the higher order terms
 of the derivative expansion in the low energy effective action.
In this sense the result of the renormalization group
 assuming the naive K\"ahler potential is an artifact
 \footnote{In general, the dynamically generated superpotential
           is non-renormalizable.}.
The value of $\alpha$
 is determined by the normalization of the kinetic term
 which is included in the unknown exact K\"ahler potential.

There are other ingredients
 to affect the value of $\alpha$ in the low energy limit.
The direct correction to the Yukawa vertex exits
 by virtue of the violation of the non-renormalization theorem
 due to the supersymmetry breaking.
The additional renormalization of the kinetic term
 could also exist due to the supersymmetry breaking.
If it could be considered that
 these effects are estimated by the running of $\alpha$
 governed by the renormalization group equation
 of the non-supersymmetric standard model
 from the supersymmetry breaking scale to the low energy,
 the value of $\alpha$ could be order of unity in the low energy limit
 (the supersymmetry breaking scale is determined in the next section).
In this paper
 we simply assume that $\alpha \simeq 1$
 is realized in the low energy limit.

In the following
 we consider the quantum correction
 based on the exact superpotential with naive K\"ahler potential
 to estimate the supersymmetry breaking effect.
All resultant corrections disappear in the supersymmetric limit.
We do not consider the quantum correction
 which includes the correction to the vertex of the exact superpotential
 by the interaction itself
 (for example,
  a diagram which includes the correction to the top-quark Yukawa vertex
  by the top-quark Yukawa interaction itself).

The tree-level superpotential of Eq.(\ref{WH-3})
 gives the supersymmetric masses of Higgs and leptoquarks
 in the low energy limit as
\begin{equation}
 W_H^{eff}
 = \beta \Lambda_3
   \left(
    \eta {\bar H}_3 H_3 + \eta_{LQ} {\bar \Phi}_3 \Phi_3
   \right),
\end{equation}
 where the dimensionless coefficient $\beta$
 is expected to be of the order of $1/4\pi$
 according to the naive dimensional analysis.
The first term gives so called $\mu$-term.
Therefore,
 $\beta \Lambda_3 \eta \simeq 100$ GeV is expected.
The mass of the leptoquarks
 must be large enough not to contradict with the direct search.
And if they couple to the particles in the first generation,
 the mass must be very heavy or the coupling must be very small
 to avoid rapid proton decay.

\subsection{Dynamical Supersymmetry Breaking}
\label{subsec:susy-breaking}

In this section
 the dynamical supersymmetry breaking
 in the system of Table \ref{susy-br}
 with the tree-level superpotential of Eq.(\ref{WS})
 and its transmission to the superpartners of known particles
 are explained.

We consider that
 the confinement scale of the $SU(2)_S$ supercolor, $\Lambda_S$,
 is larger than that of the $SU(2)_3$ hypercolor, $\Lambda_3$.
Therefore,
 the $SU(2)_3$ hypercolor interaction is weak at the scale $\Lambda_S$.
Then, the system of Table \ref{susy-br}
 is simply considered as the $SU(2)_S$ gauge theory
 with four doublets matter which interact with singlet fields
 through the superpotential of Eq.(\ref{WS}).
The confinement at the scale $\Lambda_S$
 and the gauge singlet low energy effective field
\begin{equation}
 {\hat V} \sim {\hat {\cal V}}
\end{equation}
 are expected (see Eq.(\ref{def-calV})).
Furthermore,
 the superpotential
\begin{equation}
 W^S_{dyn}
 = A \left( {\rm Pf} {\hat V} - (\beta \Lambda_S)^2 \right)
\end{equation}
 is expected to be generated dynamically,
 where $A$ is a Lagrange multiplier chiral superfield
 and $\beta \simeq 1/4\pi$ according to the naive dimensional analysis.
This superpotential
 gives the following constraint between the components
 of the matrix ${\hat V}$.
\begin{equation}
 V^2 = (\beta \Lambda_S)^2 + V'^2 + [{\tilde V}_1 {\tilde V}_2].
\label{constraint}
\end{equation}
Namely, $V$ is not the independent field.

The tree-level superpotential of Eq.(\ref{WS})
 gives the following low energy effective superpotential.
\begin{eqnarray}
 W_S^{eff}
 &=&
  \lambda \beta \Lambda_S {\rm tr}( {\hat Z} {\hat V} )
  + 4 \lambda_V \beta \Lambda_S Z V
  + M_X [X_1 X_2]
\nonumber\\
 &=&
  - 4 (\lambda - \lambda_V) \beta \Lambda_S Z V
  - 4 \lambda \beta \Lambda_S
       \left(
        Z' V' + {1 \over 2} {\tilde Z}^{\alpha i} {\tilde V}_{\alpha i}
       \right)
  + M_X [X_1 X_2].
\end{eqnarray}
By substituting the constraint of Eq.(\ref{constraint})
 to the above equation,
 we have
\begin{eqnarray}
 W_S^{eff}
 &=&
 - 4 (\lambda - \lambda_V) Z
   \left\{
    (\beta\Lambda_S)^2 + {1 \over 2} V'^2
    + {1 \over 2} [{\tilde V}_1 {\tilde V}_2]
    + {\cal O}\left( (\beta\Lambda_S)^{-2} \right)
   \right\}
\nonumber\\
&&
 - 4 \lambda \beta\Lambda_S
   \left(
    Z' V' + {1 \over 2} {\tilde Z}^{\alpha i} {\tilde V}_{\alpha i}
   \right)
 + M_X [X_1 X_2].
\label{WS-eff}
\end{eqnarray}
This shows that
 the system becomes the O'Raifeartaigh model at the low energy.
Namely, the auxiliary component of $Z$ gets vacuum expectation value of
\begin{equation}
 \langle F_Z^{\dag} \rangle
 = - \langle {{\partial W_S^{eff}} \over {\partial Z}} \rangle
 = 4 (\lambda - \lambda_S) (\beta \Lambda_S)^2,
\end{equation}
 and the supersymmetry is spontaneously broken.
The supersymmetry is dynamically broken
 due to the modification of the moduli space
 by the quantum effect\cite{IY,IT}.

The mass spectrum of this system at the low energy
 is briefly described as follows.
The second term of Eq.(\ref{WS-eff}) means that
 the pairs of $Z'$ and $V'$ and ${\tilde Z}$ and ${\tilde V}$
 have supersymmetric masses of
 $4\lambda\beta\Lambda_S$ and $m' \equiv 2\lambda\beta\Lambda_S$,
 respectively.
The supersymmetric mass of $X_1$ and $X_2$, $M_X$,
 is assumed to be of the order of $m'$
 as discussed in section \ref{subsec:contents}.
Since the auxiliary component of $Z$ has vacuum expectation value,
 the scalar components of $V'$ and ${\tilde V}$
 have soft supersymmetry breaking masses of
 $(4(\lambda-\lambda_S)\beta\Lambda_S)^2$ and
 $F \equiv 8((\lambda-\lambda_S)\beta\Lambda_S)^2$, respectively.
Note that ${\tilde V}$ couples to $SU(2)_3$ hypercolor.
The supersymmetry breaking
 is transmitted by this particle to known particles
 through the $SU(2)_3$ hypercolor interaction,
 and the masses of squarks, sleptons and gauginos are generated.
The fermion component of $Z$
 is the Nambu-Goldstone fermion
 associated with the spontaneous supersymmetry breaking.

The scalar component of $Z$ is massless,
 since it parameterizes the pseudo-flat direction
 which is common in the O'Raifeartaigh model at tree level.
If the pseudo-flat direction is lifted up
 by the effect of the non-trivial K\"ahler potential,
 it becomes massive.
If the resultant squared mass is negative,
 it obtains vacuum expectation value
 and $U(1)_R$ symmetry is spontaneously broken.
A mechanism which is proposed in Ref. \cite{Witten}
 is a candidate of the dynamics to have such vacuum expectation value.
In that mechanism
 the scale of the vacuum expectation value
 can naturally be much larger than that of the supersymmetry breaking.
But it can not be directly applied to our model.
In this paper
 we simply assume that
 $m \equiv 2(\lambda-\lambda_V) \langle A_Z \rangle \ne 0$
 and $m \gg \sqrt{F}$,
 where $A_Z$ is the scalar component of $Z$.
Therefore,
 the fields $V'$ and ${\tilde V}$ have supersymmetric masses,
 $2m$ and $m$, respectively,
 which break $U(1)_R$ symmetry.

The detailed descriptions
 on the mass spectrum of the $SU(2)_3$ hypercolored particles,
 ${\tilde Z}$ and ${\tilde V}$,
 and the transmission of the supersymmetry breaking to known particles
 are given in Ref.\cite{KO}.
In the following
 we give only a brief sketch of the transmission
 of the supersymmetry breaking to the superpartners of known particles.

The gaugino of the $SU(2)_3$ hypercolor becomes massive
 due to the one-loop diagram
 which includes the propagators of ${\tilde V}$.
\begin{equation}
 m_\lambda^{SU(2)_3} \simeq {{\alpha_H} \over {8\pi}} {F \over m},
\end{equation}
 where $\alpha_H$ is the gauge coupling
 of the $SU(2)_3$ hypercolor interaction.
The gauginos of standard model gauge groups become massive
 due to the two-loop diagram
 which includes the propagator of the $SU(2)_3$ gaugino
 and ``preons'' $C_3$ and $D_3$ in Table \ref{contents-3}.
\begin{equation}
 m_{\lambda_N}
 \simeq {{3\alpha_N} \over {8\pi}}
        \left( {{\alpha_H} \over {4\pi}} \right)^2
        {F \over m}
        \ln \left( {m \over {\Lambda_3}} \right),
\end{equation}
 where $N=1,2$ and $3$ correspond to the standard model interactions
 $U(1)_Y$, $SU(2)_L$ and $SU(3)_C$, respectively.
The scalar components of ``preons''
 $C_3$, $D_3$ and $N_3$ in Table \ref{contents-3}
 become massive due to the two-loop diagrams
 which include the propagators
 of ${\tilde V}$ and the $SU(2)_3$ gauge boson and gaugino.
\begin{equation}
 {\tilde m}_P^2
 \simeq {3 \over 2}
        \left( {{\alpha_H} \over {4\pi}} \right)^2
        \left( {F \over m} \right)^2.
\end{equation}
Note that
 these results are based on the perturbation
 on the gauge coupling of the $SU(2)_3$ hypercolor interaction.
Therefore,
 we can not use these results at low energy,
 since the hypercolor interaction becomes strong.
We assume that putting $\alpha_H/4\pi \simeq 1$ in the above formulae
 gives at least correct order of magnitude of these quantities.
Namely, we use
\begin{equation}
 m_{\lambda_N}
 \simeq {{3\alpha_N} \over {8\pi}}
        {F \over m}
        \ln \left( {m \over {\Lambda_3}} \right),
\qquad
 {\tilde m}_P^2
 \simeq {3 \over 2}
        \left( {F \over m} \right)^2
\label{soft-br-mass}
\end{equation}
 in the following.
In the next section
 we see that the condition
 $m_\lambda^{SU(2)_3}, {\tilde m}_P \ll \Lambda_3$ is satisfied,
 which is necessary to the argument in the last section.

The scalar components of the composite fields
 which belong to $M$ of Eq.(\ref{identify})
 are expected to have masses of the order of ${\tilde m}_P$.
The scalar components of the elementary particles becomes massive
 due to the one-loop diagram
 which includes massive gaugino propagator.
For example,
 the right-handed scalar bottom quark obtains the mass of
\begin{equation}
 m_{{\tilde b}_R}^2
 \simeq {{4\alpha_3} \over {3\pi}} m_{\lambda_3}^2
        \ln \left( {{\Lambda_3^2} \over {m_{\lambda_3}^2}} \right).
\end{equation}
In the next section
 we introduce the first and second generations.
Since the particles in these generations
 do not strongly couple to the supersymmetry breaking sector,
 all masses of squarks and sleptons in these generations are generated
 in the same way of the mass of the right-handed scalar bottom quark.
Therefore,
 the soft supersymmetry breaking masses of these two generations
 are highly degenerate in flavor,
 which is a good feature
 for the flavor-changing neutral current problem
 in the supersymmetric theory\cite{DN}.

\subsection{Electroweak Symmetry Breaking}
\label{subsec:ew-breaking}

In this model
 the electroweak symmetry breaking
 is triggered by the radiative correction
 due to the strong top-quark Yukawa coupling\cite{radiative}.
In the low energy limit we have the Higgs potential as
\begin{equation}
 V_{tree}
 = {1 \over 2} \left( \mu^2 + {\tilde m}_{\bar H}^2 \right) {\bar h}^2
 + {1 \over 2} \left( \mu^2 + {\tilde m}_H^2 \right) h^2
 + B {\bar h} h
 + {{g_1^2 + g_2^2} \over {32}} \left( {\bar h}^2 - h^2 \right)^2,
\end{equation}
 where ${\bar h}/\sqrt{2}$ and $h/\sqrt{2}$
 are the electrically neutral scalar components
 of ${\bar H}$ and $H$, respectively,
 $\mu=\beta\Lambda_3\eta$ is the supersymmetric mass,
 ${\tilde m}_{\bar H}$, ${\tilde m}_H$ and $B$
 are soft supersymmetry breaking masses,
 and $g_1$ and $g_2$ are the gauge couplings
 of $U(1)_Y$ and $SU(2)_L$ gauge interactions, respectively.
Since $H$ is composite, ${\tilde m}_H^2 \simeq {\tilde m}_P^2$.
Other supersymmetry breaking masses
 are generated as the quantum correction\cite{KO}.
\begin{equation}
 {\tilde m}^2_{\bar H}
 \simeq {{\alpha_1} \over {8\pi}}
 \left\{
   \mu_{LQ}^2
    \ln \left( 1 + {{{\tilde m}_P^2} \over {\mu_{LQ}^2}} \right)
 - \mu^2
    \ln \left( 1 + {{{\tilde m}_P^2} \over {\mu^2}} \right)
 + {\tilde m}_P^2
    \ln \left(
         {{\mu_{LQ}^2 + {\tilde m}_P^2} \over {\mu^2 + {\tilde m}_P^2}}
        \right)
 \right\},
\label{soft-Hbar}
\end{equation}
\begin{eqnarray}
 B &=& {{3\alpha_2} \over {4\pi}}
       {{m_{\lambda_2} \mu} \over {m_{\lambda_2}^2 - \mu^2}}
       \left\{
        m_{\lambda_2}^2
        \ln \left( {{\Lambda_3^2} \over {m_{\lambda_2}^2}} \right)
        - \mu^2
          \ln \left( {{\Lambda_3^2} \over {\mu^2}} \right)
       \right\}
\nonumber\\
   &+& {{\alpha_1} \over {2\pi}}
       {{m_{\lambda_1} \mu} \over {m_{\lambda_1}^2 - \mu^2}}
       \left\{
        m_{\lambda_1}^2
        \ln \left( {{\Lambda_3^2} \over {m_{\lambda_1}^2}} \right)
        - \mu^2
          \ln \left( {{\Lambda_3^2} \over {\mu^2}} \right)
       \right\},
\label{B-term}
\end{eqnarray}
 where $\mu_{LQ}=\beta\Lambda_3\eta_{LQ}$
 is the supersymmetric mass of the leptoquark.
Both quantities vanish in the supersymmetric limit
 (${\tilde m}_P^2, \ m_{\lambda_{1,2}} \rightarrow 0$).
The quantity ${\tilde m}_{\bar H}$ also vanishes, if $\mu_{LQ} = \mu$.
The electroweak symmetry breaking does not occur
 in this tree-level potential,
 since the relation
 $B^2 < (\mu^2 + {\tilde m}_{\bar H}^2 ) (\mu^2 + {\tilde m}_H^2 )$
 is satisfied
 (note that ${\tilde m}_{\bar H}^2 > 0$ and $(B/\mu)^2 < {\tilde m}_H^2$).

The one-loop diagrams of the top quark and the scalar top quark
 give the contribution of
\begin{equation}
 V_{1-loop}
 = {3 \over {8\pi^2}} \int_0^{\Lambda_3^2} dk^2 k^2
   \left\{
    \ln \left( 1 + {{g_t^2 h^2 / 2} \over {k^2 + m_{\tilde t}^2}} \right)
     - \ln \left( 1 + {{g_t^2 h^2 / 2} \over {k^2}} \right)
   \right\},
\end{equation}
 where $g_t$ is the Yukawa coupling of the top quark at the low energy,
 and $m_{\tilde t}^2 \simeq {\tilde m}_P^2$
 is the mass of the scalar top quark.
This contribution vanishes in the supersymmetric limit
 ($m_{\tilde t}^2 \simeq {\tilde m}_P^2 \rightarrow 0$).

The stationary conditions of whole potential
 with respect to the fields ${\bar h}$ and $h$
 give the following conditions.
\begin{equation}
 \mu^2 + {\tilde m}_{\bar H}^2
 + B \tan\beta + {1 \over 2} m_Z^2 \cos 2\beta = 0,
\label{condition-1}
\end{equation}
\begin{eqnarray}
 \mu^2 &+& {\tilde m}_H^2 + B \cot\beta - {1 \over 2} m_Z^2 \cos 2\beta
\nonumber\\
 &-& {3 \over {4\pi^2}} {{m_t^2} \over {v^2 \sin^2\beta}}
     \left\{
      \left( m_{\tilde t}^2 + m_t^2 \right)
       \ln {{m_{\tilde t}^2 + m_t^2 + \Lambda_3^2}
            \over
            {m_{\tilde t}^2 + m_t^2}}
      - m_t^2 \ln {{m_t^2 + \Lambda_3^2} \over {m_t^2}}
     \right\} = 0,
\label{condition-2}
\end{eqnarray}
 where valuables $h$ and ${\bar h}$
 are replaced by $v$ and $\beta$
 following $h = v \sin\beta$ and ${\bar h} = v \cos\beta$,
 and the constraints
 $(g_1^2+g_2^2) v^2 / 4 = m_Z^2$ and $g_t^2 h^2 / 2 = m_t^2$
 are applied.
The first condition determines $\tan\beta$,
 and the second condition determines the compositeness scale $\Lambda_3$.

We estimate the numerical values of $\tan\beta$ and $\Lambda_3$.
The soft supersymmetry breaking mass of preons, ${\tilde m}_P^2$,
 have to be large enough for heavy gauginos (see Eq.(\ref{soft-br-mass})).
We set ${\tilde m}_P^2 = (10{\rm TeV})^2$ and $m/\Lambda_3 \simeq 10^3$,
 then
\begin{eqnarray}
 m_{\lambda_1} &\simeq& 70 \ {\rm GeV}, \\
 m_{\lambda_2} &\simeq& 200 \ {\rm GeV}, \\
 m_{\lambda_3} &\simeq& 700 \ {\rm GeV}.
\end{eqnarray}
Once gaugino masses are determined,
 the value of $B$ can be determined up to logarithm
 giving the value of $\mu$.
We set $\mu=-50$ GeV and assume
 $\ln (\Lambda_3/m_{\lambda_{1,2}}) \simeq \ln (\Lambda_3/\mu) \simeq 14$,
 (this assumption will be consistently confirmed by the result),
 then
\begin{equation}
 B \simeq - 2000 \ {\rm (GeV)}^2.
\end{equation}
Moreover, we set $\mu_{LQ}=1$ TeV, then
\begin{equation}
 {\tilde m}_{\bar H}^2 \simeq (50 \ {\rm GeV})^2.
\end{equation}
By putting all these values to
 Eqs. (\ref{condition-1}) and (\ref{condition-2}),
 we have
\begin{equation}
 \tan\beta \simeq 1.5,
\end{equation}
\begin{equation}
 \Lambda_3 \simeq 10^8 \ {\rm GeV}.
\end{equation}
The value of $\Lambda_3$
 is consistent with the above assumption on the logarithmic factor,
 since $\ln (\Lambda_3/100{\rm GeV}) \simeq 14$.
We can obtain the value of the top-quark Yukawa coupling
 from the value of $\tan\beta$ as
\begin{equation}
 g_t = \sqrt{{{2 m_t^2} \over {v^2 \sin^2\beta}}} \simeq 1.2
\end{equation}
 which is consistent with the expectation
 that the coupling is of the order of unity.
The obtained value of $\Lambda_3$ means that
 the compositeness scale of the third generation is very large.

\section{Three-Generation Model}
\label{sec:three}

In this section
 we construct the three-generation model
 by introducing other two generations
 to the model described in section \ref{sec:one}.
We consider that
 the first and second generations
 have the same compositeness structure of the third generation.
The particle contents are shown in Table \ref{contents-123}.
Here,
 we have introduced two new hypercolor gauge interactions,
 $SU(2)_1$ and $SU(2)_2$, whose confinement scales
 are $\Lambda_1$ and $\Lambda_2$, respectively.
It is expected that
 all known particle contents of three generations
 with a pair of Higgs and a pair of leptoquarks in each generation
 are realized below the confinement scales
 in the same way of which is described in section \ref{subsec:confinement}.
The dynamically generated superpotential at the low energy is expected as
\begin{equation}
 W_{dyn}
  = \alpha \left(
            H_i q_i {\bar u}_i
             + \Phi_i q_i q_i
             + \Phi_i {\bar u}_i {\bar e}_i
            \right),
\label{Wdyn}
\end{equation}
 where $i=1,2$ and $3$ denotes the generation and $\alpha \simeq 4\pi$
 (see Eq.(\ref{Wdyn-3})).
Note that
 there is no generation mixing in these Yukawa couplings,
 since the confining dynamics are different in each generation.

We have also introduced three global $U(1)$ symmetries
 in each generation (see also Table \ref{susy-br}).
This $U(1)_i$ symmetry is a kind of the generation symmetry
 which forbids the mixing among generations.
Note that it distinguishes
 the preons $N_i$ which are singlet under the standard model gauge group
 from the preons $C_i$ and $D_i$
 which have the quantum number of the standard model gauge group.
This symmetry is explicitly broken
 only by the $SU(2)_i$ hypercolor gauge anomaly,
 and the breaking effect appears only through
 the dynamically generated superpotential of Eq.(\ref{Wdyn})\cite{IY2}.

The tree-level superpotential of Eq.(\ref{WH-3}) is generalized as
\begin{equation}
 W_H = \eta^i {\bar H}_i [D_i N_i]
     + \eta_{LQ}^i {\bar \Phi}_i [C_i N_i].
\end{equation}
In the low energy limit
 this superpotential gives mass terms of Higgs and leptoquarks.
\begin{equation}
 W_H^{eff}
 = \beta \left(
            \eta^i \Lambda_i {\bar H}_i H_i
          + \eta_{LQ}^i \Lambda_i {\bar \Phi}_i \Phi_i
         \right),
\end{equation}
 where $\beta \simeq 1/4\pi$.

The leptoquark in the first generation, $\Phi_1$, must be very heavy,
 otherwise proton rapidly decays by its tree-level exchange
 through the baryon number violating interaction in Eq.(\ref{Wdyn}).
It must be heavier than at least $10^{17}$ GeV
 so that the life time of proton is longer than $10^{32}$ years\cite{PDG}.
Therefore, we set the confinement scale of the first generation
 $\Lambda_1 \simeq 10^{18}$ GeV with $\eta_{LQ}^1 \simeq 1$.
On the other hand,
 the confinement scale of the third generation
 has already determined in section \ref{subsec:susy-breaking}
 as $\Lambda_3 \simeq 10^8$ GeV
 in accord with the radiative electroweak symmetry breaking.
This hierarchy among compositeness scales can be understood as follows.

We consider that
 the gauge couplings of three $SU(2)_i$ hypercolors coincide
 at very high energy (unification, for example).
Since the third generation
 strongly couples to the supersymmetry breaking sector,
 or the supersymmetry breaking sector
 includes $SU(2)_3$ hypercolored particles,
 the running of the $SU(2)_3$ gauge coupling is slower than the others,
 and the hierarchy, $\Lambda_3 \ll \Lambda_1 \simeq \Lambda_2$,
 can be realized.
In fact,
 the one-loop $\beta$-function of the $SU(2)_3$ hypercolor gauge coupling
 vanishes at the high energy, since there are 12 doublets ($N_f = 6$).

\begin{figure}
$$
\mbox{\epsfig{file=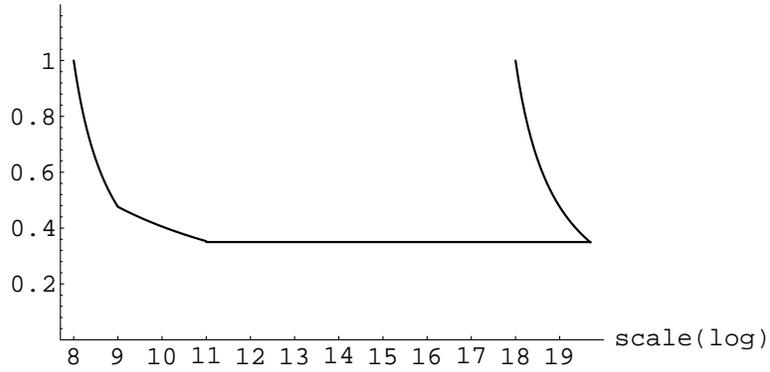,width=10cm}}
$$
\caption{Running of hypercolor gauge couplings.}
\label{running}
\end{figure}

We can understand
 from the superpotential of Eq.(\ref{WS-eff})
 how the $SU(2)_3$ hypercolored particles
 in the supersymmetry breaking sector decouple.
At the scale $m$
 the effective field ${\tilde V}$ decouples,
 where $m$ is the supersymmetric mass of ${\tilde V}$
 and we are assuming $m/\Lambda_3 \simeq 10^3$.
Furthermore, at the scale $m'$
 the field ${\tilde Z}$ decouples.
Since we are assuming $M_X \simeq m'$,
 the field $X$ also decouples at this scale.
Therefore,
 we can imagine the running of hypercolor gauge couplings
 as Fig. \ref{running}.
The ``unification scale'' $\Lambda_U$
 and the value of the ``unified gauge coupling'' $\alpha_U$
 are obtained as follows
 using the one-loop renormalization group equations.
\begin{eqnarray}
 \Lambda_U &\simeq& 5 \times 10^{19} \ {\rm GeV}, \\
 \alpha_U &\simeq& 0.35.
\end{eqnarray}
The ``unification scale'' is about the Planck scale,
 and the ``unified gauge coupling'' is rather strong.

Although we do not specify the physics at the energy scale $\Lambda_U$,
 we assume that the following kinetic-term mixing is generated.
\begin{eqnarray}
 K_{eff}
 &=& \left(
      \begin{array}{ccc}
       {\bar u}_1^{\dag} & {\bar u}_2^{\dag} & {\bar u}_3^{\dag}
      \end{array}
     \right)
     \left(
      \begin{array}{ccc}
       1 & \varepsilon & 0 \\
       \varepsilon & 1 & \varepsilon \\
       0 & \varepsilon & 1
      \end{array}
     \right)
     \left(
      \begin{array}{c}
       {\bar u}_1 \\ {\bar u}_2 \\ {\bar u}_3
      \end{array}
     \right)
\nonumber\\
 &+& \left(
      \begin{array}{ccc}
       q_1^{\dag} & q_2^{\dag} & q_3^{\dag}
      \end{array}
     \right)
     \left(
      \begin{array}{ccc}
       1 & \varepsilon & 0 \\
       \varepsilon & 1 & \varepsilon \\
       0 & \varepsilon & 1
      \end{array}
     \right)
     \left(
      \begin{array}{c}
       q_1 \\ q_2 \\ q_3
      \end{array}
     \right),
\label{kinetic-mix}
\end{eqnarray}
 where we neglect gauge fields, for simplicity.
It is not unnatural that
 the composite particles, ${\bar u}_i$ and $q_i$, have kinetic-term mixing
 at the tree level of the low energy effective Lagrangian.
The origin of the kinetic-term mixing could be considered as follows.
It can be expected that
 the physics at the scale $\Lambda_U$ equally treats each generation,
 and generates higher dimensional interaction
 in the K\"ahler potential like
\begin{equation}
 K_{\Lambda_U} = {1 \over {\Lambda_U^2}}
 \left(
  \begin{array}{ccc}
   \left[ C_1 C_1 \right]^{\dag} &
   \left[ C_2 C_2 \right]^{\dag} &
   \left[ C_3 C_3 \right]^{\dag}
  \end{array}
 \right)
 \left(
  \begin{array}{ccc}
   1 & 1 & 1 \\
   1 & 1 & 1 \\
   1 & 1 & 1
  \end{array}
 \right)
 \left(
  \begin{array}{c}
   \left[ C_1 C_1 \right] \\
   \left[ C_2 C_2 \right] \\
   \left[ C_3 C_3 \right]
  \end{array}
 \right).
\end{equation}
In the low energy limit
 this interaction gives kinetic-term mixing
\begin{equation}
 K_{\Lambda_U}^{eff} =
 \left(
  \begin{array}{ccc}
   {\bar u}_1^{\dag} &
   {\bar u}_2^{\dag} &
   {\bar u}_3^{\dag}
  \end{array}
 \right)
 {1 \over {\Lambda_U^2}}
 \left(
  \begin{array}{ccc}
   \Lambda^2 & \Lambda^2 & \Lambda\Lambda_3 \\
   \Lambda^2 & \Lambda^2 & \Lambda\Lambda_3 \\
   \Lambda_3\Lambda & \Lambda_3\Lambda & \Lambda_3^2
  \end{array}
 \right)
 \left(
  \begin{array}{c}
   {\bar u}_1 \\
   {\bar u}_2 \\
   {\bar u}_3
  \end{array}
 \right),
\end{equation}
 where $\Lambda \equiv \Lambda_1 = \Lambda_2$.
Since $\Lambda^2 / \Lambda_U^2 \simeq 0.01$
 and $\Lambda\Lambda_3 / \Lambda_U^2 \simeq 10^{-12}$,
 the kinetic-term mixing between the first and the second generations
 is relatively large,
 and the mixings between the third generation and the others
 are very small.
To have the structure of Eq.(\ref{kinetic-mix}),
 we have to include some new mechanisms to generate the large mixing
 between the second and the third generations.
An example of such mechanism has proposed in Ref.\cite{Kitazawa}.
Note that
 the kinetic-term mixings among leptoquarks and among Higgs fields
 are forbidden by the $U(1)_i$ symmetry.

We can not avoid
 to consider the hierarchical structure in the couplings
 $\eta^i$ and $\eta_{LQ}^i$.
The couplings $\eta_{LQ}^i$ must have the values of
\begin{equation}
 \eta_{LQ}^1 \simeq \eta_{LQ}^2 \simeq 1,
 \qquad
 \eta_{LQ}^3 \simeq 10^{-4}.
\end{equation}
The small value of $\eta_{LQ}^3$ is needed to have an appropriate value
 of the soft supersymmetry breaking mass ${\tilde m}^2_{\bar H}$
 for the radiative electroweak symmetry breaking
 (see Eq.(\ref{soft-Hbar})).
At present the couplings $\eta^i$ are determined as
\begin{equation}
 \eta^1 \simeq \eta^2 \simeq {\bar \eta},
 \qquad
 \eta^3 \simeq 10^{-5}.
\label{eta}
\end{equation}
The value of ${\bar \eta}$ will be determined in the next section.

The superpotential of Eq.(\ref{Wd-3})
 which connects between the supersymmetry breaking sector and remainder
 is generalized as follows due to the introduction
 of the first and second generations.
\begin{eqnarray}
 W_d &=& \kappa_{{\bar H}i} {\bar H}_i [D_3 X_1]
       + \kappa_{li} l_i [D_3 X_2]
\nonumber\\
     &+& \kappa_{{\bar \Phi}i} {\bar \Phi}_i [C_3 X_1]
       + \kappa_{{\bar d}i} {\bar d}_i [C_3 X_2]
\nonumber\\
     &+& \kappa_{{\bar \nu}i} {\bar \nu}_i [N_3 X_1].
\label{Wd}
\end{eqnarray}
Only the hypercolor-singlet fields
 in the fundamental particles of the first and second generations
 can be included.
Here, the third generation and remainder are not equally treated.

\section{Generation of the Quark Mass Hierarchy}
\label{sec:hierarchy}

\subsection{Up-type Quarks}
\label{subsec:up}

We have already shown in section \ref{subsec:ew-breaking} that
 the electroweak symmetry breaking is triggered
 by the radiatively-induced vacuum expectation values
 of the Higgs fields in the third generation, $H_3$ and ${\bar H}_3$.
The top quark becomes massive
 through the Yukawa coupling which is generated dynamically.
On the other hand,
 the Higgs fields in the first and second generations
 can not have vacuum expectation values,
 if their supersymmetric masses,
 or the coupling ${\bar \eta}$ in Eq.(\ref{eta}),
 is large enough.
Therefore,
 the up and charm quarks remain massless,
 even if there are strong Yukawa couplings
 with the Higgs fields in each generation.
In the following we describe that
 the idea which is briefly introduced in section \ref{sec:intro}
 is realized in this model.

\begin{figure}
$$
\mbox{\epsfig{file=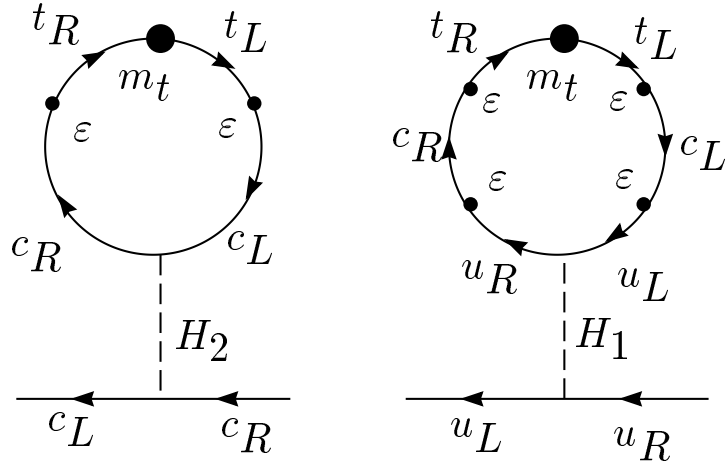,width=10cm}}
$$
\caption{The diagrams for the charm and up quarks mass generation.
         The solid and dashed line denote the fermion and boson,
         respectively.}
\label{up-type-mass}
\end{figure}

Two ingredients are needed to have the quark mass hierarchy.
One is the kinetic-term mixing among quarks,
 and the other is the strong interactions
 which are peculiar to each generation.
The kinetic-term mixing
 has already introduced in Eq.(\ref{kinetic-mix}).
The required strong interactions
 are dynamically-generated Yukawa interactions of Eq.(\ref{Wdyn}).
The up-type quarks in the first and second generations
 interact only with the Higgs field $H_1$ and $H_2$
 through the Yukawa interactions, respectively.
Then, we have the hierarchical masses of the charm and up quarks
 through the diagrams of Fig. \ref{up-type-mass}.
These tadpole diagrams
 can also be understood as the effect of the Higgs particle mixing
 which is induced by the kinetic-term mixing.
Since there is no direct transition from the up quark to the top quark,
 the mass of the up quark is suppressed by the factor $\varepsilon^2$
 in comparison with the charm quark mass.
The evaluation of the diagram gives
\begin{equation}
 m_c \simeq {{6\alpha^2} \over {16\pi^2}}
            \varepsilon^2 m_t {{{\tilde m}_P^2} \over {M_H^2}},
\qquad
 m_u \simeq \varepsilon^2 m_c,
\end{equation}
 where $\alpha \simeq 4\pi$
 and $M_H \equiv \beta \Lambda {\bar\eta}$
 with $\Lambda \equiv \Lambda_1 = \Lambda_2$
 is the common mass of the Higgs fields $H_1$ and $H_2$.
Since the quadratic divergence of these diagrams
 is canceled out in the supersymmetric limit,
 we take the mass of the scalar top quark
 $m_{\tilde t}^2 \simeq {\tilde m}_P^2$ as the ultraviolet cutoff.

If the mass $M_H$ is of the order of ${\tilde m}_P$,
 namely $M_H \simeq 10$ TeV, we have a good hierarchy
\begin{equation}
 {{m_c} \over {m_t}} \simeq {{m_u} \over {m_c}} \simeq \varepsilon^2
\end{equation}
 with $\varepsilon \simeq 0.07$, since
\begin{equation}
 {{m_c} \over {m_t}} \simeq 7 \times 10^{-3},
\qquad
 {{m_u} \over {m_c}} \simeq 4 \times 10^{-3},
\end{equation}
 where $m_t \simeq 190$ GeV, $m_c \simeq 1.4$ GeV and $m_u \simeq 5$ MeV
 at the energy scale $1$ GeV.

Taking $M_H \simeq 10$ TeV means
 a huge hierarchy in the couplings $\eta^i$ of Eq.(\ref{eta}).
Namely,
 ${\bar \eta}$ must be of the order of $10^{-13}$
 because $M_H = \beta \Lambda {\bar \eta}$
 and $\Lambda \simeq 10^{18}$ GeV.
Moreover,
 we have to note that this value of $M_H$ is marginal
 not to have the vacuum expectation values of $H_1$ and $H_2$
 due to the radiative correction by strong Yukawa couplings.

\subsection{Down-type Quarks}
\label{subsec:down}

The generation of the mass hierarchy among down-type quarks
 is not simple in comparison with that of up-type quarks.
At first,
 we show that all right-handed down-type quarks
 can couple to Higgs fields ${\bar H}_i$ in the low energy limit.

The fields $X_1$ and $X_2$
 which are included in the tree-level superpotential of Eq.(\ref{Wd})
 have the supersymmetric mass of $M_X \simeq m' \simeq 10^9$ GeV.
Therefore,
 these fields decouple before the confinement of the third generation.
Since soft supersymmetry breaking masses
 are negligible at this high energy scale
 (${\tilde m}_P \simeq m^{SU(2)_3}_\lambda \simeq 10^4$ GeV),
 we can integrate out these fields by using the conditions of
 $\partial (W_d + W_S^{eff}) / \partial X_{1,2} = 0$,
 where $W_S^{eff}$ includes the mass term of these fields
 (see Eq.(\ref{WS-eff})).
The resultant effective superpotential is
\begin{eqnarray}
 W_d' = - {1 \over {M_X}}
 &\Bigg\{&
      \kappa_{{\bar H}i} \kappa_{lj}
       {\bar H}_i \left[ D_3 D_3 \right] l_j
   +  \kappa_{{\bar H}i} \kappa_{{\bar d}j}
       {\bar H}_i \left[ D_3 C_3 \right] {\bar d}_j
\nonumber\\
  &+& \kappa_{{\bar \Phi}i} \kappa_{lj}
       {\bar \Phi}_i \left[ C_3 D_3 \right] l_j
   +  \kappa_{{\bar \Phi}i} \kappa_{{\bar d}j}
       {\bar \Phi}_i \left[ C_3 C_3 \right] {\bar d}_j
\nonumber\\
  &+& \kappa_{{\bar \nu}i} \kappa_{lj}
       {\bar \nu}_i \left[ N_3 D_3 \right] l_j
   +  \kappa_{{\bar \nu}i} \kappa_{{\bar d}j}
       {\bar \nu}_i \left[ N_3 C_3 \right] {\bar d}_j
 \Bigg\}.
\end{eqnarray}
In the low energy limit (below the confinement scale $\Lambda_3$)
 this superpotential gives the Yukawa couplings of
\begin{eqnarray}
 W_d^{eff} = - {{\beta\Lambda_3} \over {M_X}}
 &\Bigg\{&
      \kappa_{{\bar H}i} \kappa_{lj}
       {\bar H}_i {\bar e}_3 l_j
   -  \kappa_{{\bar H}i} \kappa_{{\bar d}j}
       {\bar H}_i q_3 {\bar d}_j
\nonumber\\
  &+& \kappa_{{\bar \Phi}i} \kappa_{lj}
       {\bar \Phi}_i q_3 l_j
   +  \kappa_{{\bar \Phi}i} \kappa_{{\bar d}j}
       {\bar \Phi}_i {\bar u}_3 {\bar d}_j
\nonumber\\
  &-& \kappa_{{\bar \nu}i} \kappa_{lj}
       {\bar \nu}_i H_3 l_j
   -  \kappa_{{\bar \nu}i} \kappa_{{\bar d}j}
       {\bar \nu}_i \Phi_3 {\bar d}_j
 \Bigg\},
\label{Wd-eff}
\end{eqnarray}
 where $\beta \simeq 1/4\pi$.
The second term gives the bottom quark mass
 and the mass mixings between left-handed bottom quark
 and the right-handed down and strange quarks
 through the vacuum expectation value of ${\bar H}_3$.

Here, we give some comments on the generation of the lepton mass.
The first and fifth terms of Eq.(\ref{Wd-eff})
 give the mass matrices of charged leptons and neutrinos, respectively.
But there is no physics which explains the hierarchy of these masses,
 except for that the tau lepton
 can be naturally heavier than the other charged leptons.
One interesting thing is a relation
 between the smallness of neutrino masses
 and the long life time of proton.
The coupling of the last term of Eq.(\ref{Wd-eff}) must be very small
 not to have rapid proton decay.
If we take the value of $\kappa_{{\bar \nu}i}$ very small,
 the masses of neutrinos become very small simultaneously.
The value of the coupling $\kappa_{{\bar \Phi}i}$
 also must be very small not to have rapid proton decay.

In the following
 we only consider the second term of Eq.(\ref{Wd-eff})
 to discuss the masses of down-type quarks.
Due to the vacuum expectation value
 of the scalar component of the Higgs field ${\bar H}_3$,
 we obtain the following mass matrix for down-type quarks.
\begin{equation}
 {\cal L}_d^{mass} = -
 \left(
 \begin{array}{ccc}
  {\bar d}_R & {\bar s}_R & {\bar b}_R
 \end{array}
 \right)
 M_d^{tree}
 \left(
  \begin{array}{c}
   d_L \\ s_L \\ b_L
  \end{array}
 \right) + {\rm h.c.},
\end{equation}
\begin{equation}
 M_d^{tree}
 = {{\beta\Lambda_3} \over {M_X}}
   {v \over \sqrt{2}} \cos \beta \ \kappa_{{\bar H}3}
   \left(
    \begin{array}{ccc}
     0 & 0& \kappa_{{\bar d}1} \\
     0 & 0& \kappa_{{\bar d}2} \\
     0 & 0& \kappa_{{\bar d}3}
    \end{array}
   \right),
\end{equation}
 where $d_{L,R}$, $s_{L,R}$ and $b_{L,R}$
 are the left-handed and right-handed
 down, strange and bottom fermion fields, respectively
 (The coefficient $\beta$ is different from the $\beta$ in $\cos \beta$).
The unique non-zero eigenvalue of this mass matrix is
\begin{equation}
 m_b = {{\beta\Lambda_3} \over {M_X}}
       {v \over \sqrt{2}} \cos \beta \
       \kappa_{{\bar H}3} \kappa_{{\bar d}3},
\end{equation}
 and it is the bottom quark mass.
The down and strange quarks remain massless at this level.
The value $m_b \simeq 5.7$ GeV at $1$ GeV gives
 $\kappa_{{\bar H}3} \kappa_{{\bar d}3} \simeq 7.3$.
This means that
 the couplings $\kappa$ are supposed to be of the order of unity,
 in general.
The value of the bottom-quark Yukawa coupling is
\begin{equation}
 g_b = {{\beta\Lambda_3} \over {M_X}}
       \kappa_{{\bar H}3} \kappa_{{\bar d}3}
     \simeq 6 \times 10^{-2}.
\end{equation}

\begin{figure}
$$
\mbox{\epsfig{file=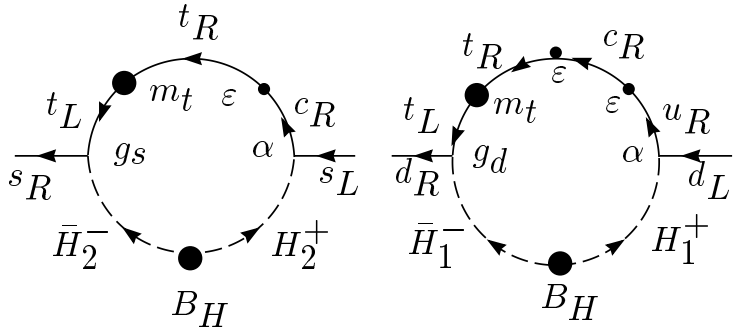,width=10cm}}
$$
\caption{The diagrams for the strange and down quarks mass generation.
         The solid and dashed line denote the fermion and boson,
         respectively.}
\label{down-type-mass}
\end{figure}

The second term of the superpotential of Eq.(\ref{Wd-eff})
 gives also the Yukawa couplings
 between two heavy Higgs, the left-handed quarks in the third generation
 and all right-handed down-type quarks.
The interplay between these Yukawa couplings
 and dynamically-generated Yukawa couplings
 gives the masses of the strange and down quarks
 through the diagrams in Fig. \ref{down-type-mass}.
In each diagram
 right couplings are dynamically-generated Yukawa couplings
 $\alpha \simeq 4\pi$,
 and left couplings are Yukawa couplings from Eq.(\ref{Wd-eff}),
\begin{equation}
 g_s = {{\beta\Lambda_3} \over {M_X}}
       \kappa_{{\bar H}2} \kappa_{{\bar d}2},
\qquad
 g_d = {{\beta\Lambda_3} \over {M_X}}
       \kappa_{{\bar H}1} \kappa_{{\bar d}1}.
\end{equation}
The soft supersymmetry breaking mass $B_H$ is obtained as
\begin{eqnarray}
 B_H
   &=& {{3\alpha_2} \over {4\pi}}
       {{m_{\lambda_2} M_H} \over {m_{\lambda_2}^2 - M_H^2}}
       \left\{
        m_{\lambda_2}^2
        \ln \left( {{\Lambda^2} \over {m_{\lambda_2}^2}} \right)
        - M_H^2
          \ln \left( {{\Lambda^2} \over {M_H^2}} \right)
       \right\}
\nonumber\\
   &&+ {{\alpha_1} \over {2\pi}}
       {{m_{\lambda_1} M_H} \over {m_{\lambda_1}^2 - M_H^2}}
       \left\{
        m_{\lambda_1}^2
        \ln \left( {{\Lambda^2} \over {m_{\lambda_1}^2}} \right)
        - M_H^2
          \ln \left( {{\Lambda^2} \over {M_H^2}} \right)
       \right\}
\nonumber\\
   &\simeq&
       {{3\alpha_2} \over {2\pi}} m_{\lambda_2} M_H
        \ln \left( {{\Lambda} \over {M_H}} \right)
    +  {{\alpha_1} \over \pi} m_{\lambda_1} M_H
        \ln \left( {{\Lambda} \over {M_H}} \right)
   \simeq (1 \ {\rm TeV})^2
\end{eqnarray}
 in exactly the same way of Eq.(\ref{B-term}).

The diagrams of Fig. \ref{down-type-mass} gives
\begin{eqnarray}
 m_s &\simeq& {{g_s \alpha} \over {16 \pi^2}}
              \varepsilon m_t {{B_H} \over {M_H^2}},
\\
 m_d &\simeq& {{g_d \alpha} \over {16 \pi^2}}
              \varepsilon^2 m_t {{B_H} \over {M_H^2}}.
\end{eqnarray}
If we naturally consider that $g_s \simeq g_d$, we have
\begin{equation}
 {{m_d} \over {m_s}} \simeq \varepsilon \simeq 0.07.
\end{equation}
Since the up quark can not directly transit to the top quark,
 the mass of the down quark is suppressed by the factor of $\varepsilon$
 than that of the strange quark.
This is a good result,
 since $m_d \simeq 10$ MeV and $m_s \simeq 200$ MeV at $1$ GeV.

The strong coupling $g_s \simeq 20$ is required
 to have the realistic value of the strange quark mass.
This value is much larger than $g_b$,
 and it corresponds to $\kappa_{{\bar H}2} \kappa_{{\bar d}2}
 = g_s M_X / \beta\Lambda_3 \simeq 3 \times 10^3$.
We have to assume the hierarchy of the order of
\begin{equation}
 \sqrt{{{\kappa_{{\bar H}2} \kappa_{{\bar d}2}}
         \over
        {\kappa_{{\bar H}3} \kappa_{{\bar d}3}}}}
 \sim 10
\end{equation}
 between the couplings in the tree level superpotential.

The origin of the bottom quark mass
 is different from the one of the down and strange quarks in this model.
The bottom quark mass
 is generated at the tree level of the low energy effective Lagrangian,
 and the others are generated at the one-loop level.
Therefore,
 it is natural that the bottom quark is heavier than that of the others,
 but the suppression due to the loop factor is too strong.
The hierarchy between the masses of the down and strange quarks
 is naturally explained by the structure of the kinetic-term mixing.

\section{Conclusion}
\label{sec:conclusion}

We have proposed a mechanism
 to have the quark mass hierarchy in the supersymmetric composite model.
The two important ingredients of the mechanism
 are the kinetic-term mixing between the composite quarks,
 which can be naturally generated due to the compositeness,
 and the strong interactions which are peculiar to each generation.
We have constructed a model in which the mechanism works.
Since we need some assumptions on the dynamics
 and the hierarchical structure between the couplings
 in the tree-level superpotential,
 the model is not completely realistic.
The another mechanism for the quark mass hierarchy
 has already been proposed 
 in the same type of composite models\cite{NS},
 and it is based on the hierarchy between many energy scales
 in the model.
A special feature of the mechanism proposed in this paper
 is its independence from the scale hierarchy in the model.

We conclude this paper
 by enumerating both the satisfactory and unsatisfactory points
 of the model.
The satisfactory points of the model are as follows.
\begin{enumerate}
\item
It is naturally understood
 why the top-quark mass is of the order of the weak scale
 and why the other quarks are much lighter than that scale.
\item
The mass hierarchy among up-type quarks
 and the one between the down and strange quarks
 can be explained only by a parameter of the kinetic-term mixing
 $\varepsilon$.
\item
The hierarchy of the compositeness scales in each generation
 can be naturally explained by considering the ``unification''
 of confining forces at the Planck scale.
\item
The representation of the fundamental particles is simple.
The representation is vector-like under each factor of the gauge group,
 and there is no particle
 which belongs to the high-dimensional representation.
\end{enumerate}

On the other hand the major problems of the model are as follows.
\begin{enumerate}
\item
There are many assumptions on the dynamics;
 the spontaneous $U(1)_R$ symmetry breaking
 and the mass generation for $X$ particle
 in the supersymmetry breaking sector,
 having the top-quark Yukawa coupling of the order of unity
 in the low energy limit,
 and the origin of the kinetic-term mixing.
\item
The hierarchy among couplings in the tree-level superpotential.
\item
The relatively light supersymmetric masses of the heavy Higgs fields
 may allow the radiative electroweak symmetry breaking
 in the first and second generations.
\item
There is no symmetry which forbids proton decay.
The difference between the baryon number and the lepton number
 is conserved, but the baryon number itself is not conserved.
The model is safe against the experimental bound,
 since we are assuming no kinetic-term mixing among charged leptons
 and small couplings in the tree-level superpotential.
 (see also the discussion in Ref. \cite{KO}).
\end{enumerate}

The mass hierarchy among the charged leptons and neutrinos
 is not explained in this model.
But it is interesting that 
 there is a relation between the smallness of neutrino masses
 and the long life time of proton.

We can discuss
 the non-trivial Cabibbo-Kobayashi-Maskawa mixing in this model.
But it is premature, because there are many problems to solve.
This model should be understood
 as a realization of the mechanism which is proposed in this paper.

\acknowledgments

This work was supported
 by U.S. Department of Energy under Contract No.~DE-FG02-92ER-40704.

\begin{table}
\begin{center}
\begin{tabular}{cccccccc}
& & $SU(2)_3$ & $SU(3)_C$ & $SU(2)_L$ & $U(1)_Y$ & $U(1)_{B-L}$ & \\
& $C_3$ & $2$ & $3$ & $1$ & $-1/3$ & $-1/6$ & \\
& $D_3$ & $2$ & $1$ & $2$ & $1/2$ & $1/2$ & \\
& $N_3$ & $2$ & $1$ & $1$ & $0$ & $-1/2$ & \\
& ${\bar d}_3$ & $1$ & $3^*$ & $1$ & $1/3$ & $-1/3$ & \\
& $l_3$ & $1$ & $1$ & $2$ & $-1/2$ & $-1$ & \\
& ${\bar \nu}_3$ & $1$ & $1$ & $1$ & $0$ & $1$ & \\
& ${\bar \Phi}_3$ & $1$ & $3^*$ & $1$ & $1/3$ & $2/3$ & \\
& ${\bar H}_3$ & $1$ & $1$ & $2$ & $-1/2$ & $0$ &
\end{tabular}
\end{center}
\caption{Particle contents of the third generation.}
\label{contents-3}
\end{table}
\newcommand{\at}[2]{$\displaystyle{#1} \atop \displaystyle{#2}$}
\begin{table}
\begin{center}
\begin{tabular}{cccccccccc}
& & $SU(2)_S$ & $SU(2)_3$ &
    $U(1)_R$ & $U(1)_{B-L}$ & $SU(2)_{global}$ & $Sp(2)_{global}$ &
    $U(1)_i$ & \\
& \at{\tilde Q}{Q} &
    \at{2}{2} & \at{2}{1} &
    \at{0}{0} & \at{0}{0} & \at{1}{2} & $4$ & \at{0}{0} & \\
& \at{\tilde Z}{Z'} &
    \at{1}{1} & \at{2^{(*)}}{1} &
    \at{2}{2} & \at{0}{0} & \at{2^{(*)}}{1} & $5$ & \at{0}{0} & \\
& $Z$ & $1$ & $1$ & $2$ & $0$ & $1$ & $1$ & 0 & \\
& $X_1$ & $1$ & $2$ & $1$ & $-1/2$ & $1$ & $1$ & 1 & \\
& $X_2$ & $1$ & $2$ & $1$ & $1/2$ & $1$ & $1$ & -1 & \\
\end{tabular}
\end{center}
\caption{Particle contents of the supersymmetry breaking sector.
         The index $i=1,2$ and $3$ denotes the generation.}
\label{susy-br}
\end{table}
\begin{table}
\begin{center}
\begin{tabular}{ccccccccc}
& & $SU(2)_i$ & $SU(3)_C$ & $SU(2)_L$ & $U(1)_Y$ & $U(1)_{B-L}$ &
    $U(1)_i$ & \\
& $C_i$ & $2$ & $3$ & $1$ & $-1/3$ & $-1/6$ & 0 & \\
& $D_i$ & $2$ & $1$ & $2$ & $1/2$ & $1/2$ & 0 & \\
& $N_i$ & $2$ & $1$ & $1$ & $0$ & $-1/2$ & 1 & \\
& ${\bar d}_i$ & $1$ & $3^*$ & $1$ & $1/3$ & $-1/3$ & 1 & \\
& $l_i$ & $1$ & $1$ & $2$ & $-1/2$ & $-1$ & 1 & \\
& ${\bar \nu}_i$ & $1$ & $1$ & $1$ & $0$ & $1$ & -1 & \\
& ${\bar \Phi}_i$ & $1$ & $3^*$ & $1$ & $1/3$ & $2/3$ & -1 & \\
& ${\bar H}_i$ & $1$ & $1$ & $2$ & $-1/2$ & $0$ & -1 &
\end{tabular}
\end{center}
\caption{Particle contents of three generations.
         The index $i=1,2$ and $3$ denotes the generation.}
\label{contents-123}
\end{table}


\end{document}